\documentclass[aps,preprint]{revtex4}
\usepackage{amsfonts}
\usepackage{amsmath}
\usepackage{amssymb}
\usepackage{graphicx}

\input{tcilatex}

\begin{document}

\title{Bernstein Waves in Symmetric and Asymmetric Pair Ions Plasma}
\author{Waseem Khan$^{1}$, Zahida Ehsan${}^{2}$ and Muddasir Ali$^{1}$}
\affiliation{$^{1}$Department of Physics, School of Natural Sciences, NUST, H-12 Campus,
Islamabad 44000, Pakistan \\
$^{2}$Department of Physics, COMSATS Institute of Information Technology,
Lahore 54000, Pakistan.}

\begin{abstract}
\end{abstract}

\begin{abstract}
\end{abstract}

\affiliation{Positive and negative ions forming so-called pair plasma differing in sign
of their charge but asymmetric in mass and temperature support a new
electrostatic mode. Bernstein mode for a pair ions and pair ions with
contribution of electrons in pair plasma both cases are investigated. By
solving the linearized Vlasov equation along Maxwell equations, a
generalized expression for the Bernstein waves is derived by employing the
Maxwell distribution function. In paper we discuss the different types of
ions Bernstein waves and comparison of the symmetry and asymmetry on these
ions Bernstein waves. We also apply the fluid limit on these Bernstein waves
and we different fluid results from kinetic theory. }
\maketitle
\date{}

\section{Introduction}

Pair plasmas have been dealt with two different ways: One is recognized as a
"symmetric" system where pair particles have the same charge, mass,
temperature, density etc., whereas in second way of treatment, the symmetry
of the pair plasma is mildly broken and the system is usually known as
"asymmetric" plasma. \ 

This asymmetry, however, brings forth new physics frontiers, as is of
interest as such plasmas can be produced in the laboratory. Whereas some
nonlinear phenomena which emerge naturally during the evolution of pair
particles may usually cause this asymmetric behavior in the experiments.
Small temperature differences in the constituent species causing asymmetries
can lead to interesting nonlinear structure formation in astrophysical
settings where one encounters e-p plasmas and in laboratory produced pair
ion plasma, whereas in the latter small contamination by a much heavier
immobile ion, or a small mass difference between the two constituent species
can also produce asymmetries \cite{20, 21, 22, 23}.

In Japan, Hatekayama and Oohara \cite{24, 25} succeeded in creating lighter
pair plasma with hydrogen; however, efforts are being made to accomplish its
improved quality. In parallel, for the theoreticians it's a challenge to
explain some of the results and some attempts have been made with kinetic
theory taking into account the boundary effects \cite{7}.

Looking at the results presented by Oohara et al., \cite{1,2}, some authors
pointed out that the produced pair-ion fullerene plasmas seem to contain
electrons as well since the ion acoustic wave observed in experiment cannot
be observed in a pure a pair ion plasma at the same temperature \cite{6}.
Later on, criteria to define pure pair ion plasma was also presented and it
was shown that the electrons are not fully filtered out and the observation
of one of the linear modes proves their presence in the system . And that
the increase in the concentration of electrons in pair-ion plasmas affects
the speed of ion acoustic wave (IAW) corresponding to the same electron
temperature\cite{4}.

Verheest et al. \cite{5} demonstrated that a strict symmetry destroys the
stationary nonlinear structures of acoustic nature and showed such nonlinear
structures can exist when there is a thermodynamic asymmetry between both
constituents.

Pair ions Bernstein Waves are similar to electron Bernstein Waves but
electrons Bernstein waves are respond at high frequency and ions Bernstein
waves responds at low frequency. Those waves are propagating at right angle
to the magnetic field and respond at low frequency are called ions Bernstein
waves. When the incoming waves having low frequency of the order of ions
cyclotron frequency. Ions respond these low frequency waves, as a result
ions Bernstein waves are produced\cite%
{baumjohann2012basic,freidberg2008plasma}

This manuscript is organized in the following manner. In Sec. II, the basic
formulation to review the non-relativistic Bernstein waves  is given and the
respective dispersion relations are obtained. Sec. III deals with the
dispersion relation for the Berensten waves in pure pair plamsa and a
special acse is dealt when electrons are also present and a quantitative
analysis is done in Sec. VI.  Finally, main findings are recapitulated in
Sec. V.

\section{Mathematical model }

To find out the dispersion relation for Bernstein waves we use the Valsov
equation along the Maxwell's equations. The linearised Vlasov equation for
uniform plasma with ambient magnetic field $\mathbf{B}_{0}$ is given as \cite%
{montgomery1964plasma,keston2003bernstein,gill2009dispersion}, 
\begin{align}
& \frac{\partial {f_{1}}}{\partial {t}}+\mathbf{v}\cdot \nabla f_{1}+\frac{q%
}{m}(\mathbf{v}\times {\mathbf{B}_{0}})\cdot \frac{\partial {f_{1}}}{%
\partial {\mathbf{v}}}  \notag \\
& =-\frac{q}{m}(\mathbf{E}_{1}+\mathbf{v}\times {\mathbf{B}_{1}})\cdot (%
\frac{\partial {f_{o}}}{\partial {\mathbf{v}}})  \tag{1}
\end{align}%
where $f_{0}$ is the equilibrium and $f_{1}$ is perturbation in the
distribution function and $q$ is the charge of the species. We find the
perturbation in the distribution function in the term of electromagnetic
fields that is given as: 
\begin{align}
f_{1}(\mathbf{r}(t),\mathbf{v}(t),t)& =-\frac{q}{m}{\int_{-\infty }^{t}}(%
\mathbf{E}_{1}(\mathbf{r}^{\prime },t^{\prime })  \notag \\
& +(\mathbf{v}^{\prime }\times (\mathbf{B}(\mathbf{r}^{\prime },t^{\prime
}))\cdot \frac{\partial {f_{0}(\mathbf{v}^{\prime })}}{\partial {\mathbf{v}}}%
dt^{\prime }  \tag{2}  \label{2}
\end{align}%
Together with the Maxwell's equations 
\begin{eqnarray}
\nabla .\mathbf{E} &=&\frac{1}{\epsilon _{0}}\sum_{s}q_{s}\int {f}d^{3}v 
\notag \\
\frac{1}{\mu _{0}}{\nabla }{\times }\mathbf{E} &=&{\epsilon _{0}}\frac{%
\partial {\mathbf{E}}}{\partial {t}}+\sum_{s}q_{s}\int {\mathbf{v}f}d^{3}v 
\notag \\
{\nabla }{\times }\mathbf{E} &=&-\frac{\partial {\mathbf{B}}}{\partial {t}} 
\TCItag{3} \\
\nabla {\cdot {\mathbf{B}}} &=&0  \notag
\end{eqnarray}%
we get, 
\begin{equation}
\overleftrightarrow{\epsilon }=1+\sum_{s}\frac{\omega _{ps}^{2}}{\omega ^{2}}%
{\sum_{n=-\infty }^{\infty }\frac{1}{n_{s}}\int }\frac{\overleftrightarrow{D}%
}{(\omega -{k_{\parallel }}{v_{\parallel }}-n\omega _{cs})}d^{3}v  \tag{4}
\end{equation}%
\begin{equation}
\overleftrightarrow{D}=\left[ {%
\begin{array}{ccc}
{v_{\perp }}(\frac{nJ_{n}{(\lambda )}}{\lambda })^{2}{P} & \frac{n}{\lambda }%
i{v_{\perp }}P{J_{n}}{(\lambda )}{J_{n}^{\prime }}{(\lambda )} & Q{v_{\perp }%
}\frac{n}{\lambda }J_{n}^{2}({\lambda }) \\ 
-i\frac{n}{\lambda }{v_{\perp }}P{J_{n}}(\lambda ){J_{n}^{\prime }}(\lambda )
& {v_{\perp }}{J_{n}^{\prime 2}(\lambda )}P & {-i}{v_{\perp }}Q{%
J_{n}(\lambda )}{J_{n}^{\prime }{(\lambda )}} \\ 
{v_{\parallel }}(\frac{n}{\lambda })J_{n}^{2}{(\lambda )}{P} & i{%
v_{\parallel }}P{J_{n}}{(\lambda )}{J_{n}^{\prime }}{(\lambda )} & Q{%
v_{\parallel }}J_{n}^{2}({\lambda })%
\end{array}%
}\right]  \tag{5}
\end{equation}%
When we use the fluid theory or cold plasma, the dielectric tensor is a
function of $\omega _{p}$ and $\omega _{c}$ only. $\overleftrightarrow{D}$
is called hot plasma dispersion tensor. The dielectric tensor is not only a
function of $\omega _{p}$ and $\omega _{c}$. It is also a function of
temperature and wave number $k$. We include the thermal motion of particles
also. In cold plasma approximation we neglect thermal motion of particles
because of which we lose some important features.

\subsection{Dielectric Tensor for an Isotropic Maxwellian Plasma}

The Maxwellian distribution is define as\cite{brambilla1998kinetic}, 
\begin{equation}
f_{0s}={n_{0s}}\left( \frac{1}{{v_{ths}}\sqrt{\pi }}\right) ^{3}\left( \exp {%
-\frac{v_{s}^{2}}{v_{ths}^{2}}}\right)  \tag{6}
\end{equation}%
\begin{equation*}
v_{ths}=\left( \frac{2kT_{s}}{m_{s}}\right) ^{\frac{1}{2}}
\end{equation*}%
$v_{ths}$ is thermal velocity of $sth$ specie. 
\begin{equation*}
v_{s}^{2}=v_{\perp }^{2}+v_{\parallel }^{2}
\end{equation*}%
\begin{equation}
P=(\omega -{k_{\parallel }{v_{\parallel }}})\frac{\partial {f_{0}}}{\partial 
{v_{\perp }}}+{k_{\parallel }{v_{\perp }}}\frac{\partial {f_{0}}}{{\partial {%
v_{\parallel }}}}  \tag{7}
\end{equation}%
\begin{equation}
Q=\frac{n\omega _{cs}v_{\parallel }}{v_{\perp }}\frac{\partial {f_{0}}}{%
\partial {v_{\perp }}}+(\omega -n\omega _{cs})\frac{\partial {f_{0}}}{{%
\partial {v_{\parallel }}}}  \tag{8}
\end{equation}%
Solving above tensor for Maxwellian distribution and we get ${\epsilon _{xx}}
$, ${\epsilon _{xz}}$ and ${\epsilon _{zz}}$ components of the tensor\cite%
{chen1984plasma}, which are used to find the dispersion relation for the
Bernstein waves. 
\begin{equation}
\epsilon _{xx}=1+\sum_{s}\frac{\omega _{ps}^{2}}{\omega ^{2}}{\xi _{0}}{%
\sum_{n=-\infty }^{\infty }}\frac{n^{2}}{b_{s}}{\exp (-b_{s})}{I_{n}(b_{s})}{%
Z(\xi _{ns})}  \tag{9}
\end{equation}%
\begin{equation}
\epsilon _{xz}=-{i}\sum_{s}\frac{\omega _{ps}^{2}}{\omega ^{2}}{\xi _{0}}{%
\sum_{n=-\infty }^{\infty }}{n}\frac{\exp (-b_{s})}{\sqrt{2b_{s}}}{%
I_{n}(b_{s})}{Z^{\prime }(\xi _{ns})}  \tag{10}
\end{equation}%
\begin{equation}
\epsilon _{zz}=1-\sum_{s}\frac{\omega _{ps}^{2}}{\omega ^{2}}{\xi _{0}}{%
\sum_{n=-\infty }^{\infty }}{\exp (-b_{s})}{I_{n}(b_{s})}\xi _{n}{Z^{\prime
}(\xi _{ns})}  \tag{11}
\end{equation}

\section{General Dispersion Relation for Bernstein Waves}

Electrostatic wave propagating at right angle to $B_{0}$ at harmonics of the
cyclotron frequency are called Bernstein wave. Poisson's equation for
electrostatic waves is written as: 
\begin{equation}
\nabla \cdot \overleftrightarrow{\epsilon }\cdot \mathbf{E}=0  \tag{12}
\end{equation}%
If we assume electrostatic perturbation such that $\mathbf{E}_{1}=-\nabla
\phi _{1}$ and consider the form of $\phi _{1}=\phi _{0}\exp [\iota (\mathbf{%
k}.\mathbf{r}-\omega {t})]$. Let $k$ is lying in x-z plane, after Fourier
transformation the Poisson's equation take the form, 
\begin{equation}
(\mathbf{k}\cdot \overleftrightarrow{\epsilon }\cdot \mathbf{k})\phi _{1}=0 
\tag{13}
\end{equation}%
$\phi _{1}{\not=0}$ so above equation is written as 
\begin{equation}
(\mathbf{k}\cdot \overleftrightarrow{\epsilon }\cdot \mathbf{k}%
)=\sum_{ij}k_{i}k_{j}\epsilon _{ij}=0  \tag{14}
\end{equation}%
The generalized expression for the Bernstein waves is written as \cite%
{chen1984plasma}, 
\begin{equation}
1+\sum_{s}\frac{k_{Ds}^{2}}{k^{2}}{\sum_{n=-\infty }^{\infty }}{\exp (-b_{s})%
}{I_{n}(b_{s})}[1+\xi _{0s}Z(\xi _{ns})]=0  \tag{15}
\end{equation}%
where s' represent the specie.

\section{Dispersion Relation for the Pair Ions Bernstein Waves }

Pair ions Bernstein Waves are similar to electron Bernstein Waves but
electrons Bernstein waves are respond at high frequency and ions Bernstein
waves responds at low frequency. \textit{Those waves are propagating at
right angle to the magnetic field and respond at low frequency are called
ions Bernstein waves.} When the incoming waves having low frequency of the
order of ions cyclotron frequency. Ions respond these low frequency waves,
as a result ions Bernstein waves are produced\cite%
{baumjohann2012basic,freidberg2008plasma}. 
\begin{equation}
\omega \approx {\omega _{p\pm }}<<\omega _{pe}  \tag{16}
\end{equation}%
Here we consider the pair plasma of positive and negative ions. So the
dispersion relation for the pair ions is written as, 
\begin{equation}
1+\frac{k_{D+}^{2}}{k^{2}}{\sum_{n=-\infty }^{\infty }}{\exp (-b_{+})}{%
I_{n}(b_{+})}[1+\xi _{0+}Z(\xi _{n+})]+\frac{k_{D-}^{2}}{k^{2}}{%
\sum_{n=-\infty }^{\infty }}{\exp (-b_{-})}{I_{n}(b_{-})}[1+\xi _{0-}Z(\xi
_{n-})]=0  \tag{17}
\end{equation}%
where $Z(\xi _{n\pm })$ is plasma dispersion function\cite%
{freidberg2008plasma}. For large value of $\xi _{n\pm }$ it is written as, 
\begin{equation}
Z(\xi _{n\pm })=\frac{-1}{\xi _{n\pm }}  \tag{18}
\end{equation}%
. where $\xi _{n\pm }$ and $\xi _{-n\pm }$ are define as 
\begin{equation}
\xi _{n\pm }=\frac{\omega -n\omega _{c\pm }}{k_{z}v_{th\pm }}  \tag{19}
\end{equation}%
and 
\begin{equation}
\xi _{-n\pm }=\frac{\omega +n\omega _{c\pm }}{k_{z}v_{th\pm }}  \tag{20}
\end{equation}%
So the dispersion relation for pair ions Bernstein waves is written as, 
\begin{equation}
1=\frac{k_{D+}^{2}}{k^{2}}{\sum_{n=1}^{\infty }}{\exp (-b_{+})}{I_{n}(b_{+})}%
\left[ \frac{2n^{2}\omega _{c+}^{2}}{(\omega ^{2}-n^{2}\omega _{c+}^{2})}%
\right] +\frac{k_{D-}^{2}}{k^{2}}{\sum_{n=1}^{\infty }}{\exp (-b_{-})}{%
I_{n}(b_{-})}\left[ \frac{2n^{2}\omega _{c-}^{2}}{(\omega ^{2}-n^{2}\omega
_{c-}^{2})}\right]  \tag{21}
\end{equation}%
Where $k_{D\pm }$ and $b_{\pm }$ are define as 
\begin{equation*}
k_{D\pm }^{2}=\frac{2\omega _{p\pm }^{2}}{v_{th\pm }^{2}}
\end{equation*}%
\begin{equation*}
b_{\pm }=\frac{k_{\perp }^{2}{v_{th\pm }^{2}}}{2\omega _{c\pm }^{2}}
\end{equation*}%
where $\omega _{p\pm }$ and $\omega _{c\pm }$ plasma and cyclotron
frequencies and ${v_{th\pm }}$ is the thermal velocity of the positive and
negative ions respectively. For symmetric case the above dispersion relation
become. 
\begin{equation}
1=2\frac{k_{D\pm }^{2}}{k^{2}}{\sum_{n=1}^{\infty }}{\exp (-b_{\pm })}{%
I_{n}(b_{\pm })}\left[ \frac{2n^{2}\omega _{c\pm }^{2}}{(\omega
^{2}-n^{2}\omega _{c\pm }^{2})}\right]  \tag{22}
\end{equation}

\subsection{Fluid Limit on the Pair Ions Bernstein Waves}

For small value of $b_{\pm }$ the modified Bessel function is written as $%
I_{n}(b_{\pm })=\frac{1}{n!}(\frac{b_{\pm }}{2})^{n}$ when $b_{\pm
}\longrightarrow {0}$ only $n=1$ term exist. So Eq.(3) is written as 
\begin{equation}
1-\frac{\omega _{p+}^{2}}{\omega ^{2}-\omega _{c+}^{2}}-\frac{\omega
_{p-}^{2}}{\omega ^{2}-\omega _{c-}^{2}}=0  \tag{23}
\end{equation}

\section{Contribution of Electrons in Pair Ions Bernstein Waves}

When we include the electrons in pair ions plasma the dispersion relation
for the pair ions Bernstein waves is modified and given as 
\begin{align}
& \frac{k_{De}^{2}}{k^{2}}{\sum_{n=-\infty }^{\infty }}{\exp (-be)}{I_{n}(be)%
}[1+\xi _{0e}Z(\xi _{ne})]+\frac{k_{D+}^{2}}{k^{2}}{\sum_{n=-\infty
}^{\infty }}{\exp (-b+)}{I_{n}(b+)}[1+\xi _{0+}Z(\xi _{n+})]+  \notag \\
\frac{k_{D-}^{2}}{k^{2}}{\sum_{n=-\infty }^{\infty }}{\exp (-b-)}{I_{n}(b-)}%
[1+\xi _{0i}Z(\xi _{n-})]& =1  \tag{24}
\end{align}

\subsection{Neutralized Pair Ions Bernstein Waves}

We consider finite $k_{z}$ such that $\frac{\omega }{k_{z}}<<v_{the}$ then $%
\xi _{ne}\longrightarrow 0$ and $Z(\xi _{ne})\approx {-2\xi _{ne}}$. For
perpendicular wavelength of the order of ion gyro radius we further have $%
b_{e}<<1$. Hence only $n=0$ term survives in the first sum. So the
dispersion relation for the Neutralized pair ions Bernstein waves is given
as, 
\begin{equation}
\frac{T_{e}}{T_{+}}{\sum_{n=1}^{\infty }}{\exp (-{b+})}{I_{n}({b+})}[\frac{%
2n^{2}\omega _{c+}^{2}}{(\omega ^{2}-n^{2}\omega _{c+}^{2})}]+\frac{T_{e}}{%
T_{-}}{\sum_{n=1}^{\infty }}{\exp (-{b-})}{I_{n}({b-})}[\frac{2n^{2}\omega
_{c-}^{2}}{(\omega ^{2}-n^{2}\omega _{c-}^{2})}]=(1+k^{2}{{\lambda _{De}^{2}}%
})  \tag{25}
\end{equation}%
For symmetric case the dispersion relation is written as, 
\begin{equation}
2\frac{T_{e}}{T_{\pm }}{\sum_{n=1}^{\infty }}{\exp (-{b\pm })}{I_{n}({b\pm })%
}\left[ \frac{2n^{2}\omega _{c\pm }^{2}}{(\omega ^{2}-n^{2}\omega _{c\pm
}^{2})}\right] =(1+k^{2}{{\lambda _{De}^{2}}})  \tag{26}
\end{equation}

\subsubsection{Fluid Limit on the Neutralized Pair Ions Bernstein Waves}

When $b_{\pm }\longrightarrow {0}$ only $n=1$ term survive, so the
dispersion relation is given as, 
\begin{equation}
1-\frac{k^{2}v_{s+}^{2}}{\omega ^{2}-\omega _{c+}^{2}}-\frac{k^{2}v_{s-}^{2}%
}{\omega ^{2}-\omega _{c-}^{2}}=0  \tag{27}
\end{equation}

\subsection{Pure Pair Ions Bernstein Waves}

In the limit of (almost) exact perpendicular propagation $\frac{\omega }{%
k_{z}}>>v_{the}$. We further assume that $b_{e}<<1$. For small value of $%
b_{e}$ the modified Bessel function is written as $I_{(n)}=\frac{1}{n!}(%
\frac{b_{e}}{2})^{n}$ when $b_{e}\longrightarrow {0}$ only $n=1$ term
survive. 
\begin{equation}
1-\frac{\omega _{pe}^{2}}{\omega ^{2}-\omega _{ce}^{2}}-\frac{2\omega
_{p+}^{2}}{k^{2}v_{th+}^{2}}{\sum_{n=1}^{\infty }}{\exp (-{b+})}{I_{n}({b+})}%
\left[ \frac{2n^{2}\omega _{c+}^{2}}{(\omega ^{2}-n^{2}\omega _{c+}^{2})}%
\right] -\frac{2\omega _{p-}^{2}}{k^{2}v_{th-}^{2}}{\sum_{n=1}^{\infty }}{%
\exp (-{b-})}{I_{n}({b-})}\left[ \frac{2n^{2}\omega _{c-}^{2}}{(\omega
^{2}-n^{2}\omega _{c-}^{2})}\right] =0  \tag{28}
\end{equation}%
In the case of ions $\omega <<\omega _{ce}$. So the dispersion relation for
the pure pair ions Bernstein waves is given as, 
\begin{align}
& 1+\frac{\omega _{pe}^{2}}{\omega _{ce}^{2}}-\frac{2\omega _{p+}^{2}}{%
k^{2}v_{th+}^{2}}{\sum_{n=1}^{\infty }}{\exp (-{b+})}{I_{n}({b+})}\left[ 
\frac{2n^{2}\omega _{c+}^{2}}{(\omega ^{2}-n^{2}\omega _{c+}^{2})}\right]  
\notag \\
-\frac{2\omega _{p-}^{2}}{k^{2}v_{th-}^{2}}{\sum_{n=1}^{\infty }}{\exp (-{b-}%
)}{I_{n}({b-})}\left[ \frac{2n^{2}\omega _{c-}^{2}}{(\omega ^{2}-n^{2}\omega
_{c-}^{2})}\right] & =0  \tag{29}
\end{align}%
For symmetric case 
\begin{equation}
1+\frac{\omega _{pe}^{2}}{\omega _{ce}^{2}}-\frac{4\omega _{p\pm }^{2}}{%
k^{2}v_{th\pm }^{2}}{\sum_{n=1}^{\infty }}{\exp (-{b}}_{{\pm }}{)}{I_{n}({b}}%
_{{\pm }}{)}\left[ \frac{2n^{2}\omega _{c\pm }^{2}}{(\omega ^{2}-n^{2}\omega
_{c\pm }^{2})}\right]   \tag{30}
\end{equation}

\section{Quantitative analysis}

For the graphical representation of the pair ions Bernstein waves we discuss
the different cases. By comparing these graph we conclude some important
results.

For the Asymmetric and Symmetric pair ions In Fig. 1, a plot of pair ions
Bernstein waves is presented by using Eq. 4. In this plot we take $%
m_{+}=m_{-}$ and $T_{-}=T_{+}$. We observe the structure of the curve is
similar to the ions Bernstein waves but due the pair ions the waves get
higher values as compare to the single specie. The dispersion curves in Fig.
2, for a pair ions Bernstein waves are shown with asymmetry in mass of ions
and temperature. Here we take $m_{+}>m_{-}$ and $T_{-}>T_{+}$. Due to
asymmetry the curve are different from the symmetric once. \ 

Now in another case, asymmetric and symmetric Pair ions with Electrons when
we include the contribution of electron with pair ions there two different
types of pair ions Bernstein waves are observed. Pair ions Bernstein waves
in which phase velocity is less than the thermal velocity of electron called
Neutralized pair ions waves. In Fig. 3 a plot for Neutralized pair ions
waves is presented by using Eq. 7, in which we first we take the symmetry in
the masses and temperature of ions but the temperature of electron is
greater than the positive and negative ions( $m_{+}=m_{-}$ and $%
T_{-}=T_{+}<T_{e}$). In Fig. 4 we present the graph for Neutralized pair
ions waves by using Eq. 6, in this plot we asymmetry in the masses and
temperature we get the curves which are different from the curves which are
plotted for symmetric once($m_{+}>m_{-}$ and $T_{-}>T_{+}<T_{e}$). When the
phase velocity is greater than the thermal velocity of electron then the
pair ions Bernstein waves are called pure pair ions Bernstein waves. The
plots in Fig. 5 for symmetric pair ions( $m_{+}=m_{-}$ and $T_{-}=T_{+}$ is
plotted by using Eq. 10 we get the get same structure of the like single
ions waves but higher values. The plots in Fig. 6 by using Eq.9 is plotted
for asymmetric pair ions(($m_{+}>m_{-}$ and $T_{-}>T_{+}$). Here we observed
that we get another harmonics which is absent in the symmetric case. In Fig.
7 and 8 we compare the results of pair ions Bernstein waves with pair ions
having contribution of electrons in it and we observe how electrons
contribution affect the pair ions Bernstein waves in symmetric and
asymmetric case respectively.

\section{Summary}

The result obtain shows that the pair ions Bernstein waves having different
propagation characteristic in symmetric and asymmetric cases. Due to
asymmetry more no of harmonics are observed. We can also observe that when
we include the electrons in pair ions plasma the curves damped more quickly
as compare to the curves in the pair ions. These results shows that more
heating is possible when electrons are a part of pair ions plasma.

\textbf{Figure Caption}

FIG. 1. Dispersion curves represents the results of pair ions having
symmetry im mass and temperature.

FIG. 2. Dispersion curves represents the results of pair ions having
asymmetry im mass and temperature.

FIG. 3. Dispersion curves represents the results of Neutralized pair ions
with electron and having symmetry im mass and temperature of the ions.

FIG. 4. Dispersion curves represents the results of Neutralized pair ions
with electron and having asymmetry im mass and temperature of the ions.

FIG. 5. Dispersion curves represents the results of pure pair ions with
electron and having symmetry im mass and tem-

perature of the ions.

FIG. 6. Dispersion curves represents the results of pure pair ions with
electron and having asymmetry im mass and tem-

perature of the ions

FIG. 7. Dispersion curves showing the comparison between symmetric pair ions
and symmetric pair ions having contri-

bution of electrons.

\end{document}